\documentclass{article}
\usepackage{amssymb}
\usepackage[mathscr]{eucal}
\usepackage{graphicx}
\usepackage[percent]{overpic} 
\usepackage{rotating} 
\usepackage[bf,centerlast]{caption}   
\usepackage{array} 
\usepackage{epsfig} 
\usepackage{psfrag} 
\newcommand{\PBS}[1]{\let\temp=\\#1\let\\=\temp} 

\psfrag{1}[r][r]{{\tiny$1$}}
\psfrag{0.8}[r][r]{{\tiny$0.8$}}
\psfrag{0.6}[r][r]{{\tiny$0.6$}}
\psfrag{0.5}[r][r]{{\tiny$0.5$}}
\psfrag{0.4}[r][r]{{\tiny$0.4$}}
\psfrag{0.3}[r][r]{{\tiny$0.3$}}
\psfrag{0.2}[r][r]{{\tiny$0.2$}}
\psfrag{0.1}[r][r]{{\tiny$0.1$}}
\psfrag{0}[r][r]{{\tiny$0$}}
\psfrag{\2610.1}[r][r]{{\tiny$-0.1$}}
\psfrag{\2610.2}[r][r]{{\tiny$-0.2$}}
\psfrag{\2610.3}[r][r]{{\tiny$-0.3$}}
\psfrag{\2610.4}[r][r]{{\tiny$-0.4$}}
\psfrag{\2610.5}[r][r]{{\tiny$-0.5$}}
\psfrag{\2610.6}[r][r]{{\tiny$-0.6$}}
\psfrag{\2610.8}[r][r]{{\tiny$-0.8$}}
\psfrag{\2611}[r][r]{{\tiny$-1$}}
\psfrag{\2611.2}[r][r]{{\tiny$-1.2$}}
\psfrag{\2612}[r][r]{{\tiny$-2$}}
\psfrag{\2614}[r][r]{{\tiny$-4$}}
\psfrag{\26110}[r][r]{{\tiny$-10$}}
\psfrag{\26120}[r][r]{{\tiny$-20$}}
\psfrag{\26130}[r][r]{{\tiny$-30$}}
\psfrag{2}[r][r]{{\tiny$2$}}
\psfrag{4}[r][r]{{\tiny$4$}}
\psfrag{5}[r][r]{{\tiny$5$}}
\psfrag{6}[r][r]{{\tiny$6$}}
\psfrag{8}[r][r]{{\tiny$8$}}
\psfrag{10}[r][r]{{\tiny$10$}}
\psfrag{12}[r][r]{{\tiny$12$}}
\psfrag{15}[r][r]{{\tiny$15$}}
\psfrag{20}[r][r]{{\tiny$20$}}
\psfrag{30}[r][r]{{\tiny$30$}}
\psfrag{ox}[t][t]{$n$}
\psfrag{oy}[r][r]{$\frac{f_{n+1}}{f_{n-1}}$}

\begin{document}

\title{The Post-Newtonian Approximation of the Rigidly Rotating Disc of Dust to
Arbitrary Order}
\author{D. Petroff and R. Meinel \\
        {\small Theoretisch-Physikalisches Institut} \\[-1.5mm]
        {\small University of Jena} \\[-1.5mm]
        {\small Max-Wien-Platz 1, 07743 Jena, Germany} \\
        {\small D.Petroff@tpi.uni-jena.de}}

\maketitle

\begin{abstract}
Using the analytic, global solution for the rigidly rotating disc of dust as a
starting point, an iteration scheme is presented for the calculation of an
arbitrary coefficient in the post-Newtonian (PN) approximation of this
solution. The coefficients were explicitly calculated up to the $12^{\rm th}$ PN
level and are listed in this paper up to the $4^{\rm th}$ PN level. The
convergence of the series is discussed and the approximation is  found to be
reliable even in highly relativistic cases. Finally, the ergospheres are
calculated at increasing orders of the approximation and for increasingly
relativistic situations.
\end{abstract}

\section{Introduction}
The study of a three-dimensional, arbitrarily rotating fluid in general
relativity cannot but rely on the use of numerical computation for its results.
An important step along the way toward a proper application of numerical
methods and toward an in-depth understanding of numerical results is the
rigorous consideration of a simpler, but related, problem. Neugebauer and
Meinel undertook this task in modelling a uniformly rotating disc of dust. The
corresponding analytic, global solution to Einstein's field equations, which
was worked out in the series of papers \cite{NM93}, \cite{NM94} and \cite{NM95}
by utilizing the `inverse scattering method' known from soliton theory, can be
expressed using hyperelliptic integrals.

In this paper, the above mentioned global solution will be used as a starting
point from which to derive an iteration scheme for the calculation of an
arbitrary term in the post-Newtonian (PN) expansion of the solution. Such an
expansion amounts to the analytic analogue of numerical work presented by
Bardeen and Wagoner \cite{BW} in a paper that handles the rotating disc in
great detail. 

Given that the global solution for the uniformly rotating disc of dust is
known, the question arises as to why one would consider the problem using the
PN approximation. The reasons are threefold. First, the complex nature of the
solution leads to fairly long computing times, particularly if the relativistic
parameter $\mu$ is to be varied. Using the PN approximation, one could speed up
calculations dramatically. Secondly, one is presented the rare opportunity to
check the accuracy of numerical work against its analytic analogue, thus
allowing one better to determine the limitations of these numerical techniques.
Lastly, since one can obtain an arbitrary term in the PN expansion, one can
look at the convergence of the series for various applications and, moreover,
see if the approximation can be used to study relativistic phenomena, where its
validity is not {\it a priori} clear.

\section{The Global Solution}
The form of the metric to be used here is taken directly from \cite{NM93} and
reads
\begin{equation}
ds^2 = e^{-2U}[e^{2k}(d\rho^2 + d\zeta^2) + \rho^2 d\varphi^2] - e^{2U}(dt + a
       d\varphi)^2
\label{eq:metric} \end{equation}
in Weyl-Lewis-Papapetrou coordinates.
\footnote{Units have been chosen in which $G$ and $c$ are equal to one, except
in eq.~(\ref{eq:mu}), where the factor $c^2$ was included to show explicitly
the relation $\mu \propto 1/c^2$ for reasons to be discussed in
section~\ref{sec:c}.}
The metric represents a stationary, axially symmetric spacetime and the three
metric functions $U, k$ and $a$ depend only on $\rho$, $\zeta$, the
relativistic parameter
\begin{equation}
\mu = \frac{2 \Omega^2 \rho_0^2 e^{-2V_0}}{c^2}
\label{eq:mu} \end{equation}
and the coordinate radius of the disc $\rho_0$. The other quantities appearing
in eq.~(\ref{eq:mu}) are the constant angular velocity  $\Omega$ and the
``surface potential'' $V_0 \equiv U(\rho\!=\!0,\zeta\!=\!0)$, which is closely
related to the redshift at infinity. The relativistic parameter $\mu$ runs from
$\mu=0$ in the Newtonian limit through to $\mu=\mu_0=4.62\ldots$ in the extreme
Kerr limit. The four-velocity of a particle in the disc has only two non-zero
components: $u^t=e^{-V_0}$ and $u^\varphi=\Omega e^{-V_0}$ (both components are
independent of the radial coordinate $\rho$). As mentioned above, the metric
depends on only two parameters, whence the profile for the surface mass density
$\sigma(\rho)$ cannot be chosen freely, but is instead automatically determined
by these parameters, i.e. a disc chosen to have the values $\mu=\tilde{\mu}$
and $\rho_0=\tilde{\rho_0}$ can have but one angular momentum $\tilde{J}$ and
but one total gravitational mass $\tilde{M}$ and the matter will needs be
distributed according to $\sigma(\rho)$ as given by eq.~(23) in \cite{NM94}.
\footnote{One could just as easily have chosen to consider a disc of mass
$M'$ and angular momentum $J'$ and would then have been
automatically led to the values for $\mu$, $\rho_0$ and $\sigma(\rho)$
corresponding to this situation.}
A specific example of a mass density profile will be presented in
section~\ref{sec:convergence}.

The vacuum field equations are equivalent to the complex Ernst
equation (see \cite{Ernst})
\begin{eqnarray}
\Re(f) \triangle f = (\nabla f)^2 \ \ \ \ \  
    \mathrm{with} \ \ \ \ \ \ f = e^{2U}+ib \label{eq:Ernst} \\ 
    \mathrm{and} \ \ b_{,\rho} = -\frac{e^{4U}}{\rho}a_{,\zeta},\ \ \ \ \  
                     b_{,\zeta} = \frac{e^{4U}}{\rho}a_{,\rho}. \label{eq:ab}
\end{eqnarray}
The operators $\triangle$ and $\nabla$ in the above equation are
three-dimensional. Taking into account boundary conditions on the disc, an
asymptotically flat solution can be found for the Ernst potential $f$, which
suffices to determine all three metric functions since $a$ can be found using
eq.~(\ref{eq:ab}) and $k$ is related to the other two functions via a line
integral as well. The potential $f$ is given by the expression
\begin{eqnarray}
\ln(f) &=& \int_0^{m_a}\frac{(X(m)-X_1)(X(m)-X_2)}{2X(m)W_1(m)}\,dm \nonumber \\
       && + \int_0^{n_b}\frac{(X(n)-X_1)(X(n)-X_2)}{2X(n)W_1(n)}\,dn \nonumber \\
       &&  - \int_{-i}^{i}\frac{H(X-X_1)(X-X_2)}{W}\,dX \label{eq:lnf} \\
       \nonumber \\
       &\equiv& I_1 + I_2 - I_3 \label{eq:I123} \end{eqnarray}
with
\begin{eqnarray}
W &=& W_1W_2, \ \ W_1 = -\sqrt{(X-\zeta/\rho_0)^2 + (\rho/\rho_0)^2},
\label{eq:W1} \\
W_2 &=& \sqrt{\mu^{-2} + (1+X^2)^2},\ \ 
H = \frac{\mathrm{arcsinh}(\mu[1+X^2])}{\pi i},
\label{eq:W2h} \\
X_1 &=& -\sqrt{\frac{i-\mu}{\mu}}\ \ {\rm and}\ \ 
X_2 = \sqrt{-\frac{i+\mu}{\mu}}.
\label{eq:X12} \end{eqnarray}
A negative sign appearing before the root of a complex quantity indicates that
the real part is to be chosen to be negative. What is meant by $X(m)$, $X(n)$,
$W_1(m)$ and $W_1(n)$ in eq.~(\ref{eq:lnf}) is that the variable substitution
\begin{equation}
X=-\sqrt{\frac{i \cosh(m)}{\mu}-1}\ \ \mathrm{or}\ \ 
X=\sqrt{\frac{-i \cosh(n)}{\mu}-1}
\label{eq:varsubmn} \end{equation}
is to be carried out. The endpoints of integration $m_a$ and $n_b$ can be
determined from the Jacobi inversion problem
\begin{equation}
\int_0^{m_a}\frac{dm}{2X(m)W_1(m)} + \int_0^{n_b}\frac{dn}{2X(n)W_1(n)} = u
\label{eq:mnu} \end{equation}
and
\begin{equation}
\int_0^{m_a}\frac{dm}{2W_1(m)} + \int_0^{n_b}\frac{dn}{2W_1(n)} = v
\label{eq:mnv} \end{equation}
with
\begin{equation}
u= \int_{-i}^i \frac{H \,dX}{W} \ \ \mathrm{and} \ \ 
v = \int_{-i}^i \frac{HX \,dX}{W}
\label{eq:uandv} \end{equation}
and where the path of integration in eq.~(\ref{eq:uandv}), as with the integral
$I_3$ in eq.~(\ref{eq:I123}), is along the imaginary axis. The above equations,
which can be found in \cite{NM95}, differ from the ones presented there in
three ways: 1)~The expression for $\ln(f)$ was manipulated algebraically,
making use of eqs.~(\ref{eq:mnu}), (\ref{eq:mnv}) and (\ref{eq:uandv}). 2)~In
each of the integrals $I_1$ and $I_2$, the aforementioned variable substitution
was carried through. 3)~The definition for $W_2$ in this paper differs from
that of \cite{NM95} by a factor $\mu$.

In discussing the Newtonian limit of the solution in \cite{NM95}, Neugebauer
and Meinel make note of the fact that the solutions can be expanded in a power
series
\begin{equation}
f= 1+ \sum_{n=1}^\infty f_n \mu^{(n+1)/2}
\label{eq:fseries} \end{equation}
such that the coefficients are elementary functions. The following section
presents a method for determining these coefficients.

\section{The Iteration Scheme}
A comparison of eqs.~(\ref{eq:varsubmn}) and (\ref{eq:X12}) shows that
\begin{equation}
X(m=0)=X_1 \ \ \ \ \mathrm{and} \ \ \ \ \ X(n=0)=X_2
\end{equation}
hold. This suggests the rearrangement of eqs.~(\ref{eq:mnu}) and (\ref{eq:mnv})
to form
\begin{eqnarray}
\int_0^{m_a}\frac{X(m)-X_2}{2X(m)W_1(m)}\,dm +
\int_0^{n_b}\frac{X(n)-X_2}{2X(n)W_1(n)}\,dn &=& v-X_2u \ ,
\label{eq:X2mn} \\
\int_0^{m_a}\frac{X(m)-X_1}{2X(m)W_1(m)}\,dm +
\int_0^{n_b}\frac{X(n)-X_1}{2X(n)W_1(n)}\,dn &=& v-X_1u,
\label{eq:X1mn} \end{eqnarray}
which serves to ``decouple'' the original equations. What is meant here by
decoupling, is that an iteration scheme for the determination of $m_a$ (or
equivalently $n_b$) as a power series in $\mu$ does not require the
simultaneous consideration of two equations in its ultimate step.
Eq.~(\ref{eq:X2mn}) yields up the ultimate term for $m_a$ and
eq.~(\ref{eq:X1mn}) for $n_b$.

Expanding the integrands in eqs.~(\ref{eq:X2mn}) and (\ref{eq:X1mn}) about the
points $m=0$ and $n=0$ respectively, one is left with trivial integrals
\begin{eqnarray}
\sum_{i=0}^\infty a_i \int_0^{m_a}m^{2i}\,dm  \ + \ 
\sum_{i=1}^\infty c_i \int_0^{n_b}n^{2i}\,dn &=& v-X_2u \label{eq:v-X2u}
\\  \nonumber \\
\sum_{i=1}^\infty d_i \int_0^{m_a}m^{2i}\,dm \ + \ 
\sum_{i=0}^\infty b_i \int_0^{n_b}n^{2i}\,dn &=& v-X_1u. \label{eq:v-X1u}
\end{eqnarray}
Making use of the fact that $m_a$ and $n_b$ are both of the order
$\mathcal{O}(\mu)$,~
\footnote{This follows from eqs.~(\ref{eq:v-X2u}) and (\ref{eq:v-X1u}) along
with the fact that the coefficients $a_i$, $b_i$, $c_i$ and $d_i$ are all of
the order ${\cal{O}}(\sqrt{\mu})$, that $u$ and $v$ are of the order
${\cal{O}}(\mu^2)$ and that $X_1$ and $X_2$ are of the order
${\cal{O}}(1/\sqrt{\mu})$.}
one can derive the following iteration formulae:
\begin{eqnarray}
m_{a_k} &=& \frac{1}{a_0} \Biggl[(v_1\mu^2+\cdots+v_k\mu^{2k}) -
          X_2(u_1\mu^2+\cdots+u_k\mu^{2k}) \nonumber \\
          && - a_1\frac{m_{a_{k-1}}^3}{3} -
          a_2\frac{m_{a_{k-2}}^5}{5} -\cdots- a_{k-1}\frac{m_{a_1}^{2k-1}}{2k-1}
          \nonumber \\
          && - c_1\frac{n_{b_{k-1}}^3}{3} -\cdots-
          c_{k-1}\frac{n_{b_1}^{2k-1}}{2k-1} \Biggr] \label{eq:ma} \\
\mathrm{and} \nonumber \\
n_{b_k} &=& \frac{1}{b_0} \Biggl[(v_1\mu^2+\cdots+v_k\mu^{2k}) -
          X_1(u_1\mu^2+\cdots+u_k\mu^{2k}) \nonumber \\
          && - b_1\frac{n_{b_{k-1}}^3}{3} -
          b_2\frac{n_{b_{k-2}}^5}{5} -\cdots- b_{k-1}\frac{n_{b_1}^{2k-1}}{2k-1}
          \nonumber \\
          && - d_1\frac{m_{a_{k-1}}^3}{3} -\cdots-
          d_{k-1}\frac{m_{a_1}^{2k-1}}{2k-1} \Biggr] \label{eq:nb}
\end{eqnarray}
where $m_{a_k}$ and $n_{b_k}$ are defined by
\begin{equation}
m_a=m_{a_k} + \mathcal{O}(\mu^{2k+1}) \ \ \ \mathrm{and} \ \ \ \
   n_b=n_{b_k} + \mathcal{O}(\mu^{2k+1})
\label{eq:maknbk} \end{equation}
and $u_j$ as well as $v_j$ are defined below in eq.~(\ref{eq:ui}).

It proves useful to augment the collection of integrals $u$ and $v$ by defining
a third integral
\begin{equation}
w = \int_{-i}^i \frac{H X^2 \,dX}{W}.
\label{eq:w} \end{equation}
Upon introducing the oblate spheroidal coordinates $\xi$ and $\eta$
\begin{equation}
\rho=\rho_0 \sqrt{(1+\xi^2)(1-\eta^2)},\ \zeta=\rho_0\xi\eta,
\ 0\leq\xi<\infty,\ -1\leq\eta\leq1
\end{equation}
one can come up with very simple, closed-form expressions for an arbitrary term
in the series expansions of $u$, $v$ and $w$ about the point $\mu=0$
\begin{equation}
u=\sum_{j=1}^\infty u_j \mu^{2j},\ \ 
v=\sum_{j=1}^\infty v_j \mu^{2j}\ \ \mathrm{and}\ \ 
w=\sum_{j=1}^\infty w_j \mu^{2j}.
\label{eq:ui} \end{equation}
These expressions are given by
\begin{eqnarray}
u_j &=& \alpha_{j-1}\int\limits_0^{\mathrm{arccot(\xi)}}\beta^{2j-1} \,dg  
  \label{eq:uvoll} \\ \nonumber \\
v_j &=& -\xi\eta\, \alpha_{j-1}\int\limits_0^{\mathrm{arccot(\xi)}}
  \tan^2(g)\beta^{2j-1}\, dg
  \label{eq:vvoll} \\ \nonumber \\
w_j &=& \alpha_{j-1}\int\limits_0^{\mathrm{arccot(\xi)}}\gamma\beta^{2j-1} \,dg
  \label{eq:wvoll} \\ \nonumber \\
\mathrm{with}  \ \ \ \ \beta &=& 1-
(1+\xi^2)(1-\eta^2)\sin^2(g)-\xi^2\eta^2\tan^2(g),
\\
\gamma &=& -
(1+\xi^2)(1-\eta^2)\sin^2(g)-\xi^2\eta^2\tan^2(g)
\\ \mathrm{and}\ \ \ \  
\alpha_j &=& \frac{(-2)^{j+1}j!}{\pi(2j+1)!!}.
\end{eqnarray}

An expansion of the integrals $I_1$ and $I_2$ yields an expression in terms of
$m_{a_k}$ and $n_{b_k}$.
\footnote{One should be careful to choose the same sign for square roots in
$I_1$ and $I_2$ as were chosen upon expanding eqs.~(\ref{eq:mnu}) and
(\ref{eq:mnv}).}
These can in turn be expressed
as a series in $\mu$ by expanding the coefficients $a_i$, $b_i$, $c_i$ and $d_i$
about the point $\mu=0$. The integral $I_3$, which can be
written in the form
\begin{equation}
I_3 = w-(X_1 + X_2)v + X_1 X_2 u,
\end{equation}
can easily be converted into a power series in $\mu$ by expanding $X_1$ and
$X_2$ about the point $\mu=0$. Thus a means of determining the coefficients
$f_n$ of eq.~(\ref{eq:fseries}) and representing them in terms of $u_j$,
$v_j$ and $w_j$ has been found.

\section{Results}
\subsection{The Ernst Potential}
\label{sec:c}
The first eight coefficients in the expansion of the Ernst Potential are
\begin{eqnarray}
f_1 &=& u_1, \label{eq:f1inuvw} \\ 
f_2 &=& -i\sqrt{2}v_1, \label{eq:f2inuvw} \\ 
f_3 &=& \frac{1}{2}u_1^{\ 2} - w_1, \label{eq:f3inuvw} \\
f_4 &=& -i \Big[ \sqrt{2}u_1v_1 + \frac{1}{\sqrt{2}}v_1 \Big],
\label{eq:f4inuvw} \\
f_5 &=& u_2 + \frac{1}{2}u_1 + \frac{1}{6}u_1^{\ 3} - u_1w_1 - v_1^{\ 2},
\label{eq:f5inuvw} \\
f_6 &=& \frac {i\sqrt{2}}{24} \Big[ - 3\,{v_{1}} - 24
\,{v_{2}} + 8\,\xi \,\eta \,{u_{1}}^{3} - 24\,{u_{1}}^{2}\,{v_{1}} \nonumber \\ && 
- 12\,{u_{1}}\,{v_{1}} + 24\,{v_{1}}\,{w_{1}} \Big], \label{eq:f6inuvw} \\
f_7 &=& - {w_{2}} - {v_{1}}^{2} + {  \frac {1}{2}} \,{w_{1}} ^{2} + {  \frac {1}{24}}
\,{u_{1}}^{4} -  {  \frac {1}{2}} \,{w_{1}}\,{u_{1}}^{2} \nonumber \\ &&
- 2\,{u_{1}} \,{v_{1}}^{2} + {  \frac {1}{2}} \,{u_{1}}^{2} + {  \frac {1}{3}}
\,{u_{1}}^{3}  - {   \frac {1}{3}} \,{u_{1}}^{3}\,\eta ^{2} \nonumber \\ &&
+ {  \frac {1 }{3}} \,{u_{1}}^{3}\,\xi ^{2} + {u_{1}}\,{u_{2}} \label{eq:f7inuvw} \\
\mathrm{and} \nonumber \\
f_8 &=&  - {  \frac {i\sqrt{2}}{48}} \Big[ - 24\,{v_{
1}}^{3} - 3\,{v_{1}} + 24\,{v_{2}} - 8\,\xi \,\eta \,{u_{1}}^{3}
 + 48\,{u_{1}}^{2}\,{v_{1}} \nonumber \\ &&
+ 30\,{u_{1}}\,{v_{1}} - 24\,{v_{1}} \,{w_{1}}  - 48\,\eta \,\xi \,{u_{1}}\,{v_{1}}^{2} - 24\,\eta 
^{2}\,{u_{1}}^{2}\,{v_{1}} \nonumber \\ && + 24\,\xi ^{2}\,{u_{1}}^{2}\,{v_{1}}
+ 48\,{v_{1}}\,{u_{2}} + 32\,{u_{1}}^{3}\,{v_{1}} - 16\,{u_{1}}
^{4}\,\xi \,\eta \nonumber \\ &&
- 48\,{w_{1}}\,{u_{1}}\,{v_{1}} + 48\,{u_{1}}\,{v_{2}} \Big].
\label{eq:f8inuvw}\end{eqnarray}

Because $u_j$, $v_j$ and $w_j$ are all real quantities, one can see immediately
that the coefficients $f_n$ are alternately real and imaginary. Since the
relativistic parameter is proportional to $1/c^2$, this series clearly exhibits
the structure of the PN approximation, in which terms occur in pairs, the first
of which is imaginary and of the order $\mathcal{O}(1/c^{2m-1}),\, m \in
{\mathbb N}$, and the second of which is real and of the order
$\mathcal{O}(1/c^{2m})$. Note that $f_1$ represents the Newtonian limit, i.e.
the first truly post-Newtonian contribution is given by $m=2$. Although the use
of the expressions $u_j$, $v_j$ and $w_j$ provides a fairly succinct notation
for the $f_n$, the coefficients still quickly become unwieldy with increasing
$n$. For example, the expression for $f_{24}$ would fill approximately 30 pages.

When the full expressions for $u_1$, $v_1$, $w_1$ and $w_2$ are substituted into
eqs.~(\ref{eq:f1inuvw}), (\ref{eq:f3inuvw}), (\ref{eq:f5inuvw}) and
(\ref{eq:f7inuvw}), one obtains, using the abbreviation ${\rm
arccot(\xi)}=\chi$, the first four coefficients in the expansion of
$e^{2U}$, $e^{2U}=1 + \sum_{n=1}^{\infty}f_{2n-1}\mu^n$,
\begin{eqnarray}
f_1 &=& - \frac {1}{\pi}\Big[ \big(3\,\xi ^{2}\,\chi  - 3\,\xi  + \chi \big)\,
\eta ^{2}  - \xi ^{2}\,\chi  + 
\xi  + \chi \, \Big], \label{eq:fullf1}\\
f_3 &=& \frac {1}{12 \pi^2} \Big[ \big(54\,\xi ^{4}\,\chi ^{2}
 + 105\,\pi \,\xi ^{4}\,\chi  - 108\,\xi ^{3}\,\chi  + 36\,\xi ^{
2}\,\chi ^{2} - 105\,\pi \,\xi ^{3} + 90\,\pi \,\xi ^{2}\,\chi 
 \nonumber \\
 & & \mbox{} + 54\,\xi ^{2} - 36\,\xi \,\chi  + 6\,\chi ^{2} - 55
\,\pi \,\xi  + 9\,\pi \,\chi \big)\eta ^{4}
+ \big( - 36\,\xi ^{4}\,\chi ^{2} \nonumber \\
 & & \mbox{} - 90\,\pi \,\xi ^{4}\,\chi  + 72\,\xi ^{3}\,\chi  + 
24\,\xi ^{2}\,\chi ^{2} + 90\,\pi \,\xi ^{3} - 72\,\pi \,\xi ^{2}
\,\chi  - 36\,\xi ^{2} - 24\,\xi \,\chi  \nonumber \\
 & & \mbox{} + 12\,\chi ^{2} + 42\,\pi \,\xi  - 6\,\pi \,\chi \big)
\eta ^{2}
+ 
6\,\xi ^{4}\,\chi ^{2} + 9\,\pi \,\xi ^{4}\,\chi  - 12\,\xi ^{3}
\,\chi  \nonumber \\
 & & \mbox{} - 12\,\xi ^{2}\,\chi ^{2} - 9\,\pi \,\xi ^{3} + 6\,
\pi \,\xi ^{2}\,\chi  + 6\,\xi ^{2} + 12\,\xi \,\chi  + 6\,\chi 
^{2} - 3\,\pi \,\xi  - 3\,\pi \,\chi \, \Big], \label{eq:fullf3}\\
f_5 &=& - \frac {1}{180 \pi^3}\Big[ \big(810\,\xi ^{6}\,\chi 
^{3} + 9225\,\pi \,\xi ^{6}\,\chi ^{2} - 3465\,\pi ^{2}\,\xi ^{6}
\,\chi  - 2430\,\xi ^{5}\,\chi ^{2} + 810\,\xi ^{4}\,\chi ^{3}
 \nonumber \\
 & & \mbox{} - 18450\,\pi \,\xi ^{5}\,\chi  + 11025\,\pi \,\xi ^{
4}\,\chi ^{2} + 3465\,\pi ^{2}\,\xi ^{5} + 2430\,\xi ^{4}\,\chi 
 - 4725\,\pi ^{2}\,\xi ^{4}\,\chi  \nonumber \\
 & & \mbox{} - 1620\,\xi ^{3}\,\chi ^{2} + 270\,\xi ^{2}\,\chi ^{
3} + 9225\,\pi \,\xi ^{4} - 15900\,\pi \,\xi ^{3}\,\chi  + 3375\,
\pi \,\xi ^{2}\,\chi ^{2} \nonumber \\
 & & \mbox{} - 810\,\xi ^{3} + 3570\,\pi ^{2}\,\xi ^{3} - 1575\,
\pi ^{2}\,\xi ^{2}\,\chi  + 810\,\xi ^{2}\,\chi  - 270\,\xi \,
\chi ^{2} + 30\,\chi ^{3} \nonumber \\
 & & \mbox{} + 4875\,\pi \,\xi ^{2} - 2670\,\pi \,\xi \,\chi  + 
135\,\pi \,\chi ^{2} + 693\,\pi ^{2}\,\xi  - 75\,\pi ^{2}\,\chi 
 + 320\,\pi \big)\eta ^{6}
\nonumber \\
 & & + \big( - 810\,
\xi ^{6}\,\chi ^{3} - 11025\,\pi \,\xi ^{6}\,\chi ^{2} + 4725\,
\pi ^{2}\,\xi ^{6}\,\chi  + 2430\,\xi ^{5}\,\chi ^{2} \nonumber \\
 & & \mbox{} + 270\,\xi ^{4}\,\chi ^{3} + 22050\,\pi \,\xi ^{5}\,
\chi  - 9405\,\pi \,\xi ^{4}\,\chi ^{2} - 4725\,\pi ^{2}\,\xi ^{5
} - 2430\,\xi ^{4}\,\chi  \nonumber \\
 & & \mbox{} + 4725\,\pi ^{2}\,\xi ^{4}\,\chi  - 540\,\xi ^{3}\,
\chi ^{2} + 450\,\xi ^{2}\,\chi ^{3} - 11025\,\pi \,\xi ^{4} + 
13560\,\pi \,\xi ^{3}\,\chi  \nonumber \\
 & & \mbox{} - 1215\,\pi \,\xi ^{2}\,\chi ^{2} + 810\,\xi ^{3} - 
3150\,\pi ^{2}\,\xi ^{3} + 675\,\pi ^{2}\,\xi ^{2}\,\chi  + 270\,
\xi ^{2}\,\chi  - 450\,\xi \,\chi ^{2} \nonumber \\
 & & \mbox{} + 90\,\chi ^{3} - 4155\,\pi \,\xi ^{2} + 690\,\pi \,
\xi \,\chi  + 45\,\pi \,\chi ^{2} - 45\,\pi ^{2}\,\xi  - 45\,\pi 
^{2}\,\chi \big)\eta ^{4}
\nonumber \\
 & & + \big(270\,\xi ^{6}\,\chi ^{3} + 
3375\,\pi \,\xi ^{6}\,\chi ^{2} - 1575\,\pi ^{2}\,\xi ^{6}\,\chi 
 - 810\,\xi ^{5}\,\chi ^{2} - 450\,\xi ^{4}\,\chi ^{3} \nonumber \\
 & & \mbox{} - 6750\,\pi \,\xi ^{5}\,\chi  + 1215\,\pi \,\xi ^{4}
\,\chi ^{2} + 1575\,\pi ^{2}\,\xi ^{5} + 810\,\xi ^{4}\,\chi  - 
675\,\pi ^{2}\,\xi ^{4}\,\chi  \nonumber \\
 & & \mbox{} + 900\,\xi ^{3}\,\chi ^{2} + 90\,\xi ^{2}\,\chi ^{3}
 + 3375\,\pi \,\xi ^{4} - 1980\,\pi \,\xi ^{3}\,\chi  - 855\,\pi 
\,\xi ^{2}\,\chi ^{2} - 270\,\xi ^{3} \nonumber \\
 & & \mbox{} + 150\,\pi ^{2}\,\xi ^{3} + 405\,\pi ^{2}\,\xi ^{2}
\,\chi  - 450\,\xi ^{2}\,\chi  - 90\,\xi \,\chi ^{2} + 90\,\chi 
^{3} + 765\,\pi \,\xi ^{2} \nonumber \\
 & & \mbox{} + 630\,\pi \,\xi \,\chi  - 135\,\pi \,\chi ^{2} - 
315\,\pi ^{2}\,\xi  + 45\,\pi ^{2}\,\chi \big)\eta ^{2}
 - 30\,\xi ^{6}\,\chi 
^{3} \nonumber \\
 & & \mbox{} - 135\,\pi \,\xi ^{6}\,\chi ^{2} + 75\,\pi ^{2}\,\xi
 ^{6}\,\chi  + 90\,\xi ^{5}\,\chi ^{2} + 90\,\xi ^{4}\,\chi ^{3}
 + 270\,\pi \,\xi ^{5}\,\chi  + 45\,\pi \,\xi ^{4}\,\chi ^{2} \nonumber \\
 & & \mbox{} - 75\,\pi ^{2}\,\xi ^{5} - 90\,\xi ^{4}\,\chi  - 45
\,\pi ^{2}\,\xi ^{4}\,\chi  - 180\,\xi ^{3}\,\chi ^{2} - 90\,\xi 
^{2}\,\chi ^{3} - 135\,\pi \,\xi ^{4} \nonumber \\
 & & \mbox{} + 135\,\pi \,\xi ^{2}\,\chi ^{2} + 30\,\xi ^{3} + 70
\,\pi ^{2}\,\xi ^{3} - 45\,\pi ^{2}\,\xi ^{2}\,\chi  + 90\,\xi ^{
2}\,\chi  + 90\,\xi \,\chi ^{2} + 30\,\chi ^{3} \nonumber \\
 & & \mbox{} - 45\,\pi \,\xi ^{2} - 90\,\pi \,\xi \,\chi  - 45\,
\pi \,\chi ^{2} + 15\,\pi ^{2}\,\xi  + 15\,\pi ^{2}\,\chi \, \big] \label{eq:fullf5}\\
{\rm and} && \nonumber \\
f_7 &=& \frac {1}{10080 \pi^4} \Big[ \big(34020\,\xi ^{8}\,\chi 
^{4} + 1908900\,\pi \,\xi ^{8}\,\chi ^{3} - 196245\,\pi ^{2}\,\xi
 ^{8}\,\chi ^{2} - 136080\,\xi ^{7}\,\chi ^{3} \nonumber \\
 & & \mbox{} + 45360\,\xi ^{6}\,\chi ^{4} - 675675\,\pi ^{3}\,\xi
 ^{8}\,\chi  - 5726700\,\pi \,\xi ^{7}\,\chi ^{2} + 3013920\,\pi 
\,\xi ^{6}\,\chi ^{3} \nonumber \\
 & & \mbox{} + 392490\,\pi ^{2}\,\xi ^{7}\,\chi  - 326340\,\pi ^{
2}\,\xi ^{6}\,\chi ^{2} + 204120\,\xi ^{6}\,\chi ^{2} - 136080\,
\xi ^{5}\,\chi ^{3} \nonumber \\
 & & \mbox{} + 22680\,\xi ^{4}\,\chi ^{4} + 675675\,\pi ^{3}\,\xi
 ^{7} + 5726700\,\pi \,\xi ^{6}\,\chi  - 1261260\,\pi ^{3}\,\xi 
^{6}\,\chi  \nonumber \\
 & & \mbox{} - 7132860\,\pi \,\xi ^{5}\,\chi ^{2} + 1544760\,\pi 
\,\xi ^{4}\,\chi ^{3} - 196245\,\pi ^{2}\,\xi ^{6} + 521850\,\pi 
^{2}\,\xi ^{5}\,\chi  \nonumber \\
 & & \mbox{} - 136080\,\xi ^{5}\,\chi  + 136080\,\xi ^{4}\,\chi 
^{2} - 179550\,\pi ^{2}\,\xi ^{4}\,\chi ^{2} - 45360\,\xi ^{3}\,
\chi ^{3} \nonumber \\
 & & \mbox{} + 5040\,\xi ^{2}\,\chi ^{4} + 1036035\,\pi ^{3}\,\xi
 ^{5} - 1908900\,\pi \,\xi ^{5} - 727650\,\pi ^{3}\,\xi ^{4}\,
\chi  \nonumber \\
 & & \mbox{} + 5223960\,\pi \,\xi ^{4}\,\chi  - 2560740\,\pi \,
\xi ^{3}\,\chi ^{2} + 272160\,\pi \,\xi ^{2}\,\chi ^{3} + 34020\,
\xi ^{4} \nonumber \\
 & & \mbox{} - 195510\,\pi ^{2}\,\xi ^{4} - 45360\,\xi ^{3}\,\chi
  + 168294\,\pi ^{2}\,\xi ^{3}\,\chi  + 22680\,\xi ^{2}\,\chi ^{2
} \nonumber \\
 & & \mbox{} - 44100\,\pi ^{2}\,\xi ^{2}\,\chi ^{2} - 5040\,\xi 
\,\chi ^{3} + 420\,\chi ^{4} + 442365\,\pi ^{3}\,\xi ^{3} - 
1105020\,\pi \,\xi ^{3} \nonumber \\
 & & \mbox{} - 132300\,\pi ^{3}\,\xi ^{2}\,\chi  + 1123500\,\pi 
\,\xi ^{2}\,\chi  - 237300\,\pi \,\xi \,\chi ^{2} + 7140\,\pi \,
\chi ^{3} \nonumber \\
 & & \mbox{} - 10549\,\pi ^{2}\,\xi ^{2} + 16758\,\pi ^{2}\,\xi 
\,\chi  - 1365\,\pi ^{2}\,\chi ^{2} + 45477\,\pi ^{3}\,\xi  - 
107520\,\pi \,\xi  \nonumber \\
 & & \mbox{} + 35840\,\pi \,\chi  - 3675\,\pi ^{3}\,\chi \big)\eta ^{
8}
+ \big( - 45360
\,\xi ^{8}\,\chi ^{4} \nonumber \\
 & & \mbox{} - 3013920\,\pi \,\xi ^{8}\,\chi ^{3} + 326340\,\pi 
^{2}\,\xi ^{8}\,\chi ^{2} + 181440\,\xi ^{7}\,\chi ^{3} + 1261260
\,\pi ^{3}\,\xi ^{8}\,\chi  \nonumber \\
 & & \mbox{} + 9041760\,\pi \,\xi ^{7}\,\chi ^{2} - 3220560\,\pi 
\,\xi ^{6}\,\chi ^{3} - 652680\,\pi ^{2}\,\xi ^{7}\,\chi  - 
219240\,\pi ^{2}\,\xi ^{6}\,\chi ^{2} \nonumber \\
 & & \mbox{} - 272160\,\xi ^{6}\,\chi ^{2} + 30240\,\xi ^{4}\,
\chi ^{4} - 1261260\,\pi ^{3}\,\xi ^{7} + 2134440\,\pi ^{3}\,\xi 
^{6}\,\chi  \nonumber \\
 & & \mbox{} - 9041760\,\pi \,\xi ^{6}\,\chi  + 7672560\,\pi \,
\xi ^{5}\,\chi ^{2} - 619920\,\pi \,\xi ^{4}\,\chi ^{3} + 326340
\,\pi ^{2}\,\xi ^{6} \nonumber \\
 & & \mbox{} + 181440\,\xi ^{5}\,\chi  + 397320\,\pi ^{2}\,\xi ^{
5}\,\chi  - 655200\,\pi ^{2}\,\xi ^{4}\,\chi ^{2} - 60480\,\xi ^{
3}\,\chi ^{3} \nonumber \\
 & & \mbox{} + 13440\,\xi ^{2}\,\chi ^{4} + 3013920\,\pi \,\xi ^{
5} - 1714020\,\pi ^{3}\,\xi ^{5} + 1058400\,\pi ^{3}\,\xi ^{4}\,
\chi  \nonumber \\
 & & \mbox{} - 5683440\,\pi \,\xi ^{4}\,\chi  + 954240\,\pi \,\xi
 ^{3}\,\chi ^{2} + 122640\,\pi \,\xi ^{2}\,\chi ^{3} - 178080\,
\pi ^{2}\,\xi ^{4} \nonumber \\
 & & \mbox{} - 45360\,\xi ^{4} + 854952\,\pi ^{2}\,\xi ^{3}\,\chi
  - 227640\,\pi ^{2}\,\xi ^{2}\,\chi ^{2} + 30240\,\xi ^{2}\,\chi
 ^{2} \nonumber \\
 & & \mbox{} - 13440\,\xi \,\chi ^{3} + 1680\,\chi ^{4} - 599172
\,\pi ^{3}\,\xi ^{3} + 1231440\,\pi \,\xi ^{3} - 370160\,\pi \,
\xi ^{2}\,\chi  \nonumber \\
 & & \mbox{} + 147000\,\pi ^{3}\,\xi ^{2}\,\chi  - 156240\,\pi \,
\xi \,\chi ^{2} + 11760\,\pi \,\chi ^{3} - 249732\,\pi ^{2}\,\xi 
^{2} \nonumber \\
 & & \mbox{} + 169848\,\pi ^{2}\,\xi \,\chi  - 10500\,\pi ^{2}\,
\chi ^{2} - 40908\,\pi ^{3}\,\xi  + 35840\,\pi \,\xi  + 35840\,
\pi \,\chi  \nonumber \\
 & & \mbox{} + 2100\,\pi ^{3}\,\chi  - 17920\,\pi ^{2} \big)\eta ^{6}
+\big(22680\,
\xi ^{8}\,\chi ^{4} + 1544760\,\pi \,\xi ^{8}\,\chi ^{3} \nonumber \\
 & & \mbox{} - 179550\,\pi ^{2}\,\xi ^{8}\,\chi ^{2} - 90720\,\xi
 ^{7}\,\chi ^{3} - 30240\,\xi ^{6}\,\chi ^{4} - 727650\,\pi ^{3}
\,\xi ^{8}\,\chi  \nonumber \\
 & & \mbox{} - 4634280\,\pi \,\xi ^{7}\,\chi ^{2} + 619920\,\pi 
\,\xi ^{6}\,\chi ^{3} + 359100\,\pi ^{2}\,\xi ^{7}\,\chi  + 
136080\,\xi ^{6}\,\chi ^{2} \nonumber \\
 & & \mbox{} + 655200\,\pi ^{2}\,\xi ^{6}\,\chi ^{2} + 90720\,\xi
 ^{5}\,\chi ^{3} - 5040\,\xi ^{4}\,\chi ^{4} + 727650\,\pi ^{3}\,
\xi ^{7} \nonumber \\
 & & \mbox{} + 4634280\,\pi \,\xi ^{6}\,\chi  - 1058400\,\pi ^{3}
\,\xi ^{6}\,\chi  - 1662360\,\pi \,\xi ^{5}\,\chi ^{2} - 594720\,
\pi \,\xi ^{4}\,\chi ^{3} \nonumber \\
 & & \mbox{} - 179550\,\pi ^{2}\,\xi ^{6} - 1077300\,\pi ^{2}\,
\xi ^{5}\,\chi  - 90720\,\xi ^{5}\,\chi  - 90720\,\xi ^{4}\,\chi 
^{2} \nonumber \\
 & & \mbox{} + 779940\,\pi ^{2}\,\xi ^{4}\,\chi ^{2} + 10080\,\xi
 ^{3}\,\chi ^{3} + 10080\,\xi ^{2}\,\chi ^{4} + 815850\,\pi ^{3}
\,\xi ^{5} \nonumber \\
 & & \mbox{} - 1544760\,\pi \,\xi ^{5} - 396900\,\pi ^{3}\,\xi ^{
4}\,\chi  + 1464960\,\pi \,\xi ^{4}\,\chi  + 884520\,\pi \,\xi ^{
3}\,\chi ^{2} \nonumber \\
 & & \mbox{} - 196560\,\pi \,\xi ^{2}\,\chi ^{3} + 422100\,\pi ^{
2}\,\xi ^{4} + 22680\,\xi ^{4} + 30240\,\xi ^{3}\,\chi  \nonumber \\
 & & \mbox{} - 886620\,\pi ^{2}\,\xi ^{3}\,\chi  + 146160\,\pi ^{
2}\,\xi ^{2}\,\chi ^{2} - 5040\,\xi ^{2}\,\chi ^{2} - 10080\,\xi 
\,\chi ^{3} + 2520\,\chi ^{4} \nonumber \\
 & & \mbox{} + 189630\,\pi ^{3}\,\xi ^{3} - 422520\,\pi \,\xi ^{3
} - 25200\,\pi ^{3}\,\xi ^{2}\,\chi  - 289800\,\pi \,\xi ^{2}\,
\chi  \nonumber \\
 & & \mbox{} + 130200\,\pi \,\xi \,\chi ^{2} - 2520\,\pi \,\chi 
^{3} + 204330\,\pi ^{2}\,\xi ^{2} - 65100\,\pi ^{2}\,\xi \,\chi 
 - 630\,\pi ^{2}\,\chi ^{2} \nonumber \\
 & & \mbox{} + 630\,\pi ^{3}\,\xi  + 630\,\pi ^{3}\,\chi \big)\eta ^{
4}
+ \big( - 5040\,
\xi ^{8}\,\chi ^{4} - 272160\,\pi \,\xi ^{8}\,\chi ^{3} \nonumber \\
 & & \mbox{} + 44100\,\pi ^{2}\,\xi ^{8}\,\chi ^{2} + 20160\,\xi 
^{7}\,\chi ^{3} + 13440\,\xi ^{6}\,\chi ^{4} + 132300\,\pi ^{3}\,
\xi ^{8}\,\chi  \nonumber \\
 & & \mbox{} + 816480\,\pi \,\xi ^{7}\,\chi ^{2} + 122640\,\pi \,
\xi ^{6}\,\chi ^{3} - 88200\,\pi ^{2}\,\xi ^{7}\,\chi  - 227640\,
\pi ^{2}\,\xi ^{6}\,\chi ^{2} \nonumber \\
 & & \mbox{} - 30240\,\xi ^{6}\,\chi ^{2} - 40320\,\xi ^{5}\,\chi
 ^{3} - 10080\,\xi ^{4}\,\chi ^{4} - 132300\,\pi ^{3}\,\xi ^{7}
 \nonumber \\
 & & \mbox{} + 147000\,\pi ^{3}\,\xi ^{6}\,\chi  - 816480\,\pi \,
\xi ^{6}\,\chi  - 186480\,\pi \,\xi ^{5}\,\chi ^{2} + 196560\,\pi
 \,\xi ^{4}\,\chi ^{3} \nonumber \\
 & & \mbox{} + 44100\,\pi ^{2}\,\xi ^{6} + 20160\,\xi ^{5}\,\chi 
 + 367080\,\pi ^{2}\,\xi ^{5}\,\chi  - 146160\,\pi ^{2}\,\xi ^{4}
\,\chi ^{2} \nonumber \\
 & & \mbox{} + 40320\,\xi ^{4}\,\chi ^{2} + 20160\,\xi ^{3}\,\chi
 ^{3} + 272160\,\pi \,\xi ^{5} - 102900\,\pi ^{3}\,\xi ^{5} + 
5040\,\pi \,\xi ^{4}\,\chi  \nonumber \\
 & & \mbox{} + 25200\,\pi ^{3}\,\xi ^{4}\,\chi  - 302400\,\pi \,
\xi ^{3}\,\chi ^{2} - 35280\,\pi \,\xi ^{2}\,\chi ^{3} - 139440\,
\pi ^{2}\,\xi ^{4} \nonumber \\
 & & \mbox{} - 5040\,\xi ^{4} - 13440\,\xi ^{3}\,\chi  + 140840\,
\pi ^{2}\,\xi ^{3}\,\chi  - 10080\,\xi ^{2}\,\chi ^{2} + 22680\,
\pi ^{2}\,\xi ^{2}\,\chi ^{2} \nonumber \\
 & & \mbox{} + 1680\,\chi ^{4} - 2660\,\pi ^{3}\,\xi ^{3} + 58800
\,\pi \,\xi ^{3} + 105840\,\pi \,\xi ^{2}\,\chi  - 2520\,\pi ^{3}
\,\xi ^{2}\,\chi  \nonumber \\
 & & \mbox{} + 35280\,\pi \,\xi \,\chi ^{2} - 11760\,\pi \,\chi 
^{3} - 28980\,\pi ^{2}\,\xi ^{2} - 24360\,\pi ^{2}\,\xi \,\chi 
 + 4620\,\pi ^{2}\,\chi ^{2} \nonumber \\
 & & \mbox{} + 420\,\pi ^{3}\,\xi  + 420\,\pi ^{3}\,\chi \big)\eta ^{
2}
+ 420\,\xi 
^{8}\,\chi ^{4} + 7140\,\pi \,\xi ^{8}\,\chi ^{3} \nonumber \\
 & & \mbox{} - 1365\,\pi ^{2}\,\xi ^{8}\,\chi ^{2} - 1680\,\xi ^{
7}\,\chi ^{3} - 1680\,\xi ^{6}\,\chi ^{4} - 3675\,\pi ^{3}\,\xi 
^{8}\,\chi  - 21420\,\pi \,\xi ^{7}\,\chi ^{2} \nonumber \\
 & & \mbox{} - 11760\,\pi \,\xi ^{6}\,\chi ^{3} + 2730\,\pi ^{2}
\,\xi ^{7}\,\chi  + 2520\,\xi ^{6}\,\chi ^{2} + 10500\,\pi ^{2}\,
\xi ^{6}\,\chi ^{2} \nonumber \\
 & & \mbox{} + 5040\,\xi ^{5}\,\chi ^{3} + 2520\,\xi ^{4}\,\chi 
^{4} + 3675\,\pi ^{3}\,\xi ^{7} + 21420\,\pi \,\xi ^{6}\,\chi  - 
2100\,\pi ^{3}\,\xi ^{6}\,\chi  \nonumber \\
 & & \mbox{} + 18900\,\pi \,\xi ^{5}\,\chi ^{2} - 2520\,\pi \,\xi
 ^{4}\,\chi ^{3} - 1365\,\pi ^{2}\,\xi ^{6} - 16310\,\pi ^{2}\,
\xi ^{5}\,\chi  - 1680\,\xi ^{5}\,\chi  \nonumber \\
 & & \mbox{} - 5040\,\xi ^{4}\,\chi ^{2} - 630\,\pi ^{2}\,\xi ^{4
}\,\chi ^{2} - 5040\,\xi ^{3}\,\chi ^{3} - 1680\,\xi ^{2}\,\chi 
^{4} + 875\,\pi ^{3}\,\xi ^{5} \nonumber \\
 & & \mbox{} - 7140\,\pi \,\xi ^{5} + 630\,\pi ^{3}\,\xi ^{4}\,
\chi  - 2520\,\pi \,\xi ^{4}\,\chi  + 16380\,\pi \,\xi ^{3}\,\chi
 ^{2} + 11760\,\pi \,\xi ^{2}\,\chi ^{3} \nonumber \\
 & & \mbox{} + 5810\,\pi ^{2}\,\xi ^{4} + 420\,\xi ^{4} + 1680\,
\xi ^{3}\,\chi  + 1190\,\pi ^{2}\,\xi ^{3}\,\chi  - 4620\,\pi ^{2
}\,\xi ^{2}\,\chi ^{2} \nonumber \\
 & & \mbox{} + 2520\,\xi ^{2}\,\chi ^{2} + 1680\,\xi \,\chi ^{3}
 + 420\,\chi ^{4} - 595\,\pi ^{3}\,\xi ^{3} - 4620\,\pi \,\xi ^{3
} - 420\,\pi ^{3}\,\xi ^{2}\,\chi  \nonumber \\
 & & \mbox{} - 13860\,\pi \,\xi ^{2}\,\chi  - 13860\,\pi \,\xi \,
\chi ^{2} - 4620\,\pi \,\chi ^{3} + 1155\,\pi ^{2}\,\xi ^{2} + 
2310\,\pi ^{2}\,\xi \,\chi  \nonumber \\
 & & \mbox{} + 1155\,\pi ^{2}\,\chi ^{2} + 525\,\pi ^{3}\,\xi  + 
525\,\pi ^{3}\,\chi \, \Big]. \label{eq:fullf7}
\end{eqnarray}
Calculating the appropriate line integrals, one can determine the remaining
metric functions $a$ and $k$. Using the expansions
\begin{equation}
a=\sum_{n=1}^{\infty}a_{2n}\mu^{(2n+1)/2}\ \ {\rm and}\ \ 
e^{2k}=1+ \sum_{n=1}^{\infty}K_{2n-1}\mu^n,
\end{equation}
one finds the expressions
\begin{eqnarray}
a_2 &=& \frac {\sqrt{2}\rho_0}{4 \pi} \Big[1- \eta^2\Big] \,\Big[ \big(15
\,\xi ^{4}\,\chi  - 15\,\xi ^{3} + 18\,\xi ^{2}\,\chi  - 13\,\xi 
 + 3\,\chi \big)\,\eta ^{2} \nonumber \\
 & & \mbox{}  - 3\,\xi ^{4}\,\chi  + 3\,\xi ^{3
} - 2\,\xi ^{2}\,\chi  + \xi  + \chi \Big], \label{eq:fulla2}\\
a_4 &=& \frac {\sqrt{2}\rho_0}{72 \pi^2} \Big[1-\eta^2 \Big] \Big[ \big(180
\,\xi ^{6}\,\chi ^{2} - 360\,\xi ^{5}\,\chi  + 252\,\xi ^{4}\,
\chi ^{2} + 180\,\xi ^{4} - 384\,\xi ^{3}\,\chi  + 108\,\xi ^{2}
\,\chi ^{2} \nonumber \\
 & & \mbox{} + 132\,\xi ^{2} - 24\,\xi \,\chi  + 36\,\chi ^{2} - 
64 \big)\eta ^{4}
+ \big( - 72\,\xi ^{6}\,\chi ^{2} + 144\,\xi ^{5}
\,\chi  \nonumber \\
 & & \mbox{} + 216\,\xi ^{4}\,\chi ^{2} + 135\,\pi \,\xi ^{4}\,
\chi  - 72\,\xi ^{4} + 360\,\xi ^{2}\,\chi ^{2} - 135\,\pi \,\xi 
^{3} + 162\,\pi \,\xi ^{2}\,\chi  \nonumber \\
 & & \mbox{} - 216\,\xi ^{2} - 144\,\xi \,\chi  + 72\,\chi ^{2}
 - 117\,\pi \,\xi  + 27\,\pi \,\chi  - 64 \big)\eta ^{2}
 \nonumber \\
 & & + 36\,\xi ^{6}\,\chi ^{2} - 72\,\xi ^{5}\,\chi  - 36\,\xi ^{4}
\,\chi ^{2} - 27\,\pi \,\xi ^{4}\,\chi  + 36\,\xi ^{4} - 36\,\xi 
^{2}\,\chi ^{2} + 27\,\pi \,\xi ^{3} \nonumber \\
 & & \mbox{} - 18\,\pi \,\xi ^{2}\,\chi  + 36\,\xi ^{2} + 72\,\xi
 \,\chi  + 36\,\chi ^{2} + 9\,\pi \,\xi  + 9\,\pi \,\chi  - 64 \Big], \label{eq:fulla4}\\
a_6 &=& \frac {\sqrt{2}\rho_0}{1440 \pi^3}\Big[1-\eta^2 \Big] \Big[ \big(
12960\,\xi ^{8}\,\chi ^{3} + 7875\,\pi \,\xi ^{8}\,\chi ^{2} - 
45045\,\pi ^{2}\,\xi ^{8}\,\chi  - 38880\,\xi ^{7}\,\chi ^{2} \nonumber \\
 & & \mbox{} + 25920\,\xi ^{6}\,\chi ^{3} - 15750\,\pi \,\xi ^{7}
\,\chi  + 15300\,\pi \,\xi ^{6}\,\chi ^{2} + 45045\,\pi ^{2}\,\xi
 ^{7} + 38880\,\xi ^{6}\,\chi  \nonumber \\
 & & \mbox{} - 97020\,\pi ^{2}\,\xi ^{6}\,\chi  - 64800\,\xi ^{5}
\,\chi ^{2} + 17280\,\xi ^{4}\,\chi ^{3} + 7875\,\pi \,\xi ^{6}
 - 25350\,\pi \,\xi ^{5}\,\chi  \nonumber \\
 & & \mbox{} + 9450\,\pi \,\xi ^{4}\,\chi ^{2} - 12960\,\xi ^{5}
 + 82005\,\pi ^{2}\,\xi ^{5} + 51840\,\xi ^{4}\,\chi  - 66150\,
\pi ^{2}\,\xi ^{4}\,\chi  \nonumber \\
 & & \mbox{} - 30240\,\xi ^{3}\,\chi ^{2} + 4800\,\xi ^{2}\,\chi 
^{3} + 10050\,\pi \,\xi ^{4} - 11850\,\pi \,\xi ^{3}\,\chi  + 
1620\,\pi \,\xi ^{2}\,\chi ^{2} \nonumber \\
 & & \mbox{} - 12960\,\xi ^{3} + 42819\,\pi ^{2}\,\xi ^{3} + 
12960\,\xi ^{2}\,\chi  - 14700\,\pi ^{2}\,\xi ^{2}\,\chi  - 4320
\,\xi \,\chi ^{2} \nonumber \\
 & & \mbox{} + 480\,\chi ^{3} + 3275\,\pi \,\xi ^{2} - 2250\,\pi 
\,\xi \,\chi  - 405\,\pi \,\chi ^{2} + 5619\,\pi ^{2}\,\xi  - 525
\,\pi ^{2}\,\chi  \nonumber \\
 & & \mbox{} + 1280\,\pi \big)\eta ^{6}
+ \big( - 12960
\,\xi ^{8}\,\chi ^{3} - 7425\,\pi \,\xi ^{8}\,\chi ^{2} \nonumber \\
 & & \mbox{} + 51975\,\pi ^{2}\,\xi ^{8}\,\chi  + 38880\,\xi ^{7}
\,\chi ^{2} - 8640\,\xi ^{6}\,\chi ^{3} + 14850\,\pi \,\xi ^{7}\,
\chi  - 8100\,\pi \,\xi ^{6}\,\chi ^{2} \nonumber \\
 & & \mbox{} - 51975\,\pi ^{2}\,\xi ^{7} - 38880\,\xi ^{6}\,\chi 
 + 94500\,\pi ^{2}\,\xi ^{6}\,\chi  + 30240\,\xi ^{5}\,\chi ^{2}
 + 11520\,\xi ^{4}\,\chi ^{3} \nonumber \\
 & & \mbox{} - 7425\,\pi \,\xi ^{6} + 11250\,\pi \,\xi ^{5}\,\chi
  - 630\,\pi \,\xi ^{4}\,\chi ^{2} + 12960\,\xi ^{5} - 77175\,\pi
 ^{2}\,\xi ^{5} \nonumber \\
 & & \mbox{} - 34560\,\xi ^{4}\,\chi  + 47250\,\pi ^{2}\,\xi ^{4}
\,\chi  - 15840\,\xi ^{3}\,\chi ^{2} + 8640\,\xi ^{2}\,\chi ^{3}
 - 3150\,\pi \,\xi ^{4} \nonumber \\
 & & \mbox{} - 4530\,\pi \,\xi ^{3}\,\chi  + 540\,\pi \,\xi ^{2}
\,\chi ^{2} + 12960\,\xi ^{3} - 26145\,\pi ^{2}\,\xi ^{3} + 4320
\,\xi ^{2}\,\chi  \nonumber \\
 & & \mbox{} + 4500\,\pi ^{2}\,\xi ^{2}\,\chi  - 7200\,\xi \,\chi
 ^{2} + 1440\,\chi ^{3} + 4335\,\pi \,\xi ^{2} - 930\,\pi \,\xi 
\,\chi  + 495\,\pi \,\chi ^{2} \nonumber \\
 & & \mbox{} - 225\,\pi ^{2}\,\xi  - 225\,\pi ^{2}\,\chi  - 640\,
\pi \big)\eta ^{4}
+ \big(4320\,\xi ^{8}\,\chi ^{3} \nonumber \\
 & & \mbox{} + 2025\,\pi \,\xi ^{8}\,\chi ^{2} - 14175\,\pi ^{2}
\,\xi ^{8}\,\chi  - 12960\,\xi ^{7}\,\chi ^{2} - 2880\,\xi ^{6}\,
\chi ^{3} - 4050\,\pi \,\xi ^{7}\,\chi  \nonumber \\
 & & \mbox{} + 1980\,\pi \,\xi ^{6}\,\chi ^{2} + 14175\,\pi ^{2}
\,\xi ^{7} + 12960\,\xi ^{6}\,\chi  - 18900\,\pi ^{2}\,\xi ^{6}\,
\chi  + 1440\,\xi ^{5}\,\chi ^{2} \nonumber \\
 & & \mbox{} - 5760\,\xi ^{4}\,\chi ^{3} + 2025\,\pi \,\xi ^{6}
 - 2610\,\pi \,\xi ^{5}\,\chi  + 4590\,\pi \,\xi ^{4}\,\chi ^{2}
 + 14175\,\pi ^{2}\,\xi ^{5} \nonumber \\
 & & \mbox{} - 4320\,\xi ^{5} - 3375\,\pi ^{2}\,\xi ^{4}\,\chi 
 + 5760\,\xi ^{4}\,\chi  + 12960\,\xi ^{3}\,\chi ^{2} + 2880\,\xi
 ^{2}\,\chi ^{3} \nonumber \\
 & & \mbox{} + 630\,\pi \,\xi ^{4} - 990\,\pi \,\xi ^{3}\,\chi 
 + 5580\,\pi \,\xi ^{2}\,\chi ^{2} - 4320\,\xi ^{3} - 90\,\pi ^{2
}\,\xi ^{3} \nonumber \\
 & & \mbox{} + 1350\,\pi ^{2}\,\xi ^{2}\,\chi  - 7200\,\xi ^{2}\,
\chi  - 1440\,\xi \,\chi ^{2} + 1440\,\chi ^{3} - 3375\,\pi \,\xi
 ^{2} \nonumber \\
 & & \mbox{} - 2430\,\pi \,\xi \,\chi  + 945\,\pi \,\chi ^{2} - 
720\,\pi ^{2}\,\xi  - 640\,\pi \big)\eta ^{2}
\nonumber \\
 & &  - 480\,\xi ^{8}\,\chi ^{3} + 405\,\pi \,\xi ^{8}\,\chi ^{2}
 + 525\,\pi ^{2}\,\xi ^{8}\,\chi  + 1440\,\xi ^{7}\,\chi ^{2} + 
960\,\xi ^{6}\,\chi ^{3} \nonumber \\
 & & \mbox{} - 810\,\pi \,\xi ^{7}\,\chi  + 900\,\pi \,\xi ^{6}\,
\chi ^{2} - 525\,\pi ^{2}\,\xi ^{7} - 1440\,\xi ^{6}\,\chi  + 300
\,\pi ^{2}\,\xi ^{6}\,\chi  \nonumber \\
 & & \mbox{} - 1440\,\xi ^{5}\,\chi ^{2} + 405\,\pi \,\xi ^{6} - 
1530\,\pi \,\xi ^{5}\,\chi  - 450\,\pi \,\xi ^{4}\,\chi ^{2} - 
125\,\pi ^{2}\,\xi ^{5} + 480\,\xi ^{5} \nonumber \\
 & & \mbox{} - 225\,\pi ^{2}\,\xi ^{4}\,\chi  - 1440\,\xi ^{3}\,
\chi ^{2} - 960\,\xi ^{2}\,\chi ^{3} + 630\,\pi \,\xi ^{4} + 90\,
\pi \,\xi ^{3}\,\chi  \nonumber \\
 & & \mbox{} - 540\,\pi \,\xi ^{2}\,\chi ^{2} + 220\,\pi ^{2}\,
\xi ^{3} + 480\,\xi ^{3} + 1440\,\xi ^{2}\,\chi  - 30\,\pi ^{2}\,
\xi ^{2}\,\chi  + 1440\,\xi \,\chi ^{2} \nonumber \\
 & & \mbox{} + 480\,\chi ^{3} + 405\,\pi \,\xi ^{2} + 810\,\pi \,
\xi \,\chi  + 405\,\pi \,\chi ^{2} - 30\,\pi ^{2}\,\xi  - 30\,\pi
 ^{2}\,\chi  - 640\,\pi \Big], \label{eq:fulla6} \\
K_1 &=& 0, \label{eq:fullk1} \\
K_3 &=& - \frac {1}{2 \pi^2} \Big[1-\eta^2 \Big] \Big[
\big(9\,\xi ^{4}\,\chi ^{2} - 18\,\xi ^{3}\,\chi  + 10\,\xi ^{2}\,
\chi ^{2} + 9\,\xi ^{2} - 14\,\xi \,\chi  + \chi ^{2} + 4 \big)\,\eta 
^{2}
\nonumber \\ &&
- \xi ^{4}\,\chi ^{2} + 2\,\xi ^{3}\,\chi  - 2\,\xi ^{2}
\,\chi ^{2} - \xi ^{2} - 2\,\xi \,\chi  - \chi ^{2} + 4 \Big], \label{eq:fullk3} \\
K_5 &=& - \frac{1}{18\pi^2}\Big[1- \eta^2 \Big] \Big[ \big(45\,\xi ^{6}\,\chi ^{2
} - 90\,\xi ^{5}\,\chi  + 63\,\xi ^{4}\,\chi ^{2} + 45\,\xi ^{4}
 - 96\,\xi ^{3}\,\chi  + 27\,\xi ^{2}\,\chi ^{2} 
 \nonumber \\
 & & \mbox{} + 33\,\xi ^{2} - 6\,\xi \,\chi  + 9\,\chi ^{2} - 16\big)\eta ^{4}
+ \big( - 18\,\xi 
^{6}\,\chi ^{2} + 36\,\xi ^{5}\,\chi  + 54\,\xi ^{4}\,\chi ^{2}
 - 18\,\xi ^{4} \nonumber \\
 & & \mbox{} + 90\,\xi ^{2}\,\chi ^{2} - 54\,\xi ^{2} - 36\,\xi 
\,\chi  + 18\,\chi ^{2} - 16)\eta ^{2}
 \nonumber \\
 & & + 9\,\xi ^{6}\,\chi ^{2} - 18\,\xi ^{5}\,\chi  - 9\,\xi ^{4}\,
\chi ^{2} + 9\,\xi ^{4} - 9\,\xi ^{2}\,\chi ^{2} + 9\,\xi ^{2} + 
18\,\xi \,\chi  + 9\,\chi ^{2} - 16 \Big] \label{eq:fullk5} \\
{\rm and} && \nonumber \\
K_7 &=& - \frac {1}{720 \pi^4} \Big[1-\eta^2 \Big] \Big[ \big(7290\,\xi ^{8}\,\chi 
^{4} - 7245\,\pi ^{2}\,\xi ^{8}\,\chi ^{2} - 29160\,\xi ^{7}\,
\chi ^{3} + 16200\,\xi ^{6}\,\chi ^{4} \nonumber \\
 & & \mbox{} + 14490\,\pi ^{2}\,\xi ^{7}\,\chi  - 16020\,\pi ^{2}
\,\xi ^{6}\,\chi ^{2} + 43740\,\xi ^{6}\,\chi ^{2} - 55080\,\xi 
^{5}\,\chi ^{3} \nonumber \\
 & & \mbox{} + 10620\,\xi ^{4}\,\chi ^{4} - 7245\,\pi ^{2}\,\xi 
^{6} + 27210\,\pi ^{2}\,\xi ^{5}\,\chi  - 29160\,\xi ^{5}\,\chi 
 - 10350\,\pi ^{2}\,\xi ^{4}\,\chi ^{2} \nonumber \\
 & & \mbox{} + 68040\,\xi ^{4}\,\chi ^{2} - 28440\,\xi ^{3}\,\chi
 ^{3} + 1800\,\xi ^{2}\,\chi ^{4} + 7290\,\xi ^{4} - 11190\,\pi 
^{2}\,\xi ^{4} \nonumber \\
 & & \mbox{} + 14934\,\pi ^{2}\,\xi ^{3}\,\chi  - 35640\,\xi ^{3}
\,\chi  - 1620\,\pi ^{2}\,\xi ^{2}\,\chi ^{2} + 26460\,\xi ^{2}\,
\chi ^{2} - 2520\,\xi \,\chi ^{3} \nonumber \\
 & & \mbox{} + 90\,\chi ^{4} + 6480\,\xi ^{2} - 5389\,\pi ^{2}\,
\xi ^{2} + 1734\,\pi ^{2}\,\xi \,\chi  - 10080\,\xi \,\chi  - 45
\,\pi ^{2}\,\chi ^{2} \nonumber \\
 & & \mbox{} + 720\,\chi ^{2} - 256\,\pi ^{2} + 1440 \big)\eta ^{6}
+ \big( - 8910\,
\xi ^{8}\,\chi ^{4} + 8775\,\pi ^{2}\,\xi ^{8}\,\chi ^{2} \nonumber \\
 & & \mbox{} + 35640\,\xi ^{7}\,\chi ^{3} - 21240\,\xi ^{6}\,\chi
 ^{4} - 17550\,\pi ^{2}\,\xi ^{7}\,\chi  + 22500\,\pi ^{2}\,\xi 
^{6}\,\chi ^{2} \nonumber \\
 & & \mbox{} - 53460\,\xi ^{6}\,\chi ^{2} + 64440\,\xi ^{5}\,\chi
 ^{3} - 16020\,\xi ^{4}\,\chi ^{4} + 8775\,\pi ^{2}\,\xi ^{6} - 
29070\,\pi ^{2}\,\xi ^{5}\,\chi  \nonumber \\
 & & \mbox{} + 35640\,\xi ^{5}\,\chi  + 17730\,\pi ^{2}\,\xi ^{4}
\,\chi ^{2} - 65880\,\xi ^{4}\,\chi ^{2} + 33480\,\xi ^{3}\,\chi 
^{3} - 3960\,\xi ^{2}\,\chi ^{4} \nonumber \\
 & & \mbox{} - 8910\,\xi ^{4} + 6570\,\pi ^{2}\,\xi ^{4} - 13890
\,\pi ^{2}\,\xi ^{3}\,\chi  + 23400\,\xi ^{3}\,\chi  + 4500\,\pi 
^{2}\,\xi ^{2}\,\chi ^{2} \nonumber \\
 & & \mbox{} - 17460\,\xi ^{2}\,\chi ^{2} + 4680\,\xi \,\chi ^{3}
 - 270\,\chi ^{4} - 720\,\xi ^{2} + 495\,\pi ^{2}\,\xi ^{2} - 930
\,\pi ^{2}\,\xi \,\chi  \nonumber \\
 & & \mbox{} - 1440\,\xi \,\chi  + 495\,\pi ^{2}\,\chi ^{2} - 720
\,\chi ^{2} - 896\,\pi ^{2} + 1440 \big)\eta ^{4}
\nonumber \\
 & & + \big( 1710\,\xi ^{8}\,\chi ^{4} - 1575\,\pi ^{2}\,\xi ^{8}\,\chi 
^{2} - 6840\,\xi ^{7}\,\chi ^{3} + 5400\,\xi ^{6}\,\chi ^{4} + 
3150\,\pi ^{2}\,\xi ^{7}\,\chi  \nonumber \\
 & & \mbox{} + 10260\,\xi ^{6}\,\chi ^{2} - 5580\,\pi ^{2}\,\xi 
^{6}\,\chi ^{2} - 9720\,\xi ^{5}\,\chi ^{3} + 5940\,\xi ^{4}\,
\chi ^{4} - 1575\,\pi ^{2}\,\xi ^{6} \nonumber \\
 & & \mbox{} + 3390\,\pi ^{2}\,\xi ^{5}\,\chi  - 6840\,\xi ^{5}\,
\chi  - 3240\,\xi ^{4}\,\chi ^{2} - 2970\,\pi ^{2}\,\xi ^{4}\,
\chi ^{2} - 4680\,\xi ^{3}\,\chi ^{3} \nonumber \\
 & & \mbox{} + 2520\,\xi ^{2}\,\chi ^{4} + 1710\,\xi ^{4} + 2190
\,\pi ^{2}\,\xi ^{4} + 14040\,\xi ^{3}\,\chi  + 1410\,\pi ^{2}\,
\xi ^{3}\,\chi  \nonumber \\
 & & \mbox{} + 1620\,\pi ^{2}\,\xi ^{2}\,\chi ^{2} - 9900\,\xi ^{
2}\,\chi ^{2} - 1800\,\xi \,\chi ^{3} + 270\,\chi ^{4} - 6480\,
\xi ^{2} - 855\,\pi ^{2}\,\xi ^{2} \nonumber \\
 & & \mbox{} + 10080\,\xi \,\chi  - 270\,\pi ^{2}\,\xi \,\chi  + 
585\,\pi ^{2}\,\chi ^{2} - 720\,\chi ^{2} - 1440 - 896\,\pi ^{2} \big)
\eta ^{2}
\nonumber \\ &&
 - 90\,\xi ^{8}\,
\chi ^{4} + 45\,\pi ^{2}\,\xi ^{8}\,\chi ^{2} + 360\,\xi ^{7}\,
\chi ^{3} - 360\,\xi ^{6}\,\chi ^{4} - 90\,\pi ^{2}\,\xi ^{7}\,
\chi  \nonumber \\
 & & \mbox{} - 540\,\xi ^{6}\,\chi ^{2} + 540\,\pi ^{2}\,\xi ^{6}
\,\chi ^{2} + 360\,\xi ^{5}\,\chi ^{3} - 540\,\xi ^{4}\,\chi ^{4}
 + 45\,\pi ^{2}\,\xi ^{6} - 570\,\pi ^{2}\,\xi ^{5}\,\chi  \nonumber \\
 & & \mbox{} + 360\,\xi ^{5}\,\chi  + 1080\,\xi ^{4}\,\chi ^{2}
 - 90\,\pi ^{2}\,\xi ^{4}\,\chi ^{2} - 360\,\xi ^{3}\,\chi ^{3}
 - 360\,\xi ^{2}\,\chi ^{4} - 90\,\xi ^{4} \nonumber \\
 & & \mbox{} + 30\,\pi ^{2}\,\xi ^{4} - 1800\,\xi ^{3}\,\chi  - 
150\,\pi ^{2}\,\xi ^{3}\,\chi  - 180\,\pi ^{2}\,\xi ^{2}\,\chi ^{
2} + 900\,\xi ^{2}\,\chi ^{2} - 360\,\xi \,\chi ^{3} \nonumber \\
 & & \mbox{} - 90\,\chi ^{4} + 720\,\xi ^{2} + 405\,\pi ^{2}\,\xi
 ^{2} + 1440\,\xi \,\chi  + 810\,\pi ^{2}\,\xi \,\chi  + 405\,\pi
 ^{2}\,\chi ^{2} + 720\,\chi ^{2} \nonumber \\
 & & \mbox{} - 1440 - 896\,\pi ^{2} \Big]. \label{eq:fullk7}
\end{eqnarray}
Eqs.~(\ref{eq:fullf1})-(\ref{eq:fullk7}) provide the full metric up to and
including the order ${\cal O}(\mu^4)$.

Making use of the metric functions, one can furthermore find the series
expansion for other quantities of interest:
\begin{eqnarray}
e^{2V_0} &=& 1 - \mu  + \frac {1}{2} \,\mu ^{2} - 
\frac {16}{9 \pi^2} \,\mu ^{3} +
\left(\frac{16}{9\pi^2} - \frac{1}{8}  \right)\, \mu^4
+ {\cal O}(\mu ^{5}), \\
\Omega \rho_0 &=& \frac {\sqrt{2}}{2} \,\mu^{1/2} - 
\frac {\sqrt{2}}{4} \,\mu ^{3/2} + 
\frac {\sqrt{2}}{16} \,\mu ^{5/2} \nonumber \\ && + 
\frac {\sqrt{2}}{2} \,\left( - 
\frac {8}{9 \pi^2} + 
\frac {1}{16} \right) \,\mu ^{7/2} + {\cal O}(\mu ^{9/2}) \\
{\rm and} && \nonumber \\
\sigma_P/\Omega &=&  \frac {\sqrt{2}\,\eta }{\pi 
^{2}}\, \mu^{1/2}  + \bigg( -\frac {7 \sqrt{2}\,\eta ^{3}}{12 \pi^2} \, 
  +  \frac {\sqrt{2}\,\eta }{4 \pi ^{2}} \bigg)\,\mu 
^{3/2} \nonumber \\
 & & \mbox{} + \bigg( 
 \frac {163 \sqrt{2}\,\eta ^{5}}{480 \pi ^{2}}  - 
{ \frac {7 \sqrt{2}\,
\eta ^{3}}{48 \pi ^{2}}}  - 
{ \frac {3 \sqrt{2}\,\eta }{32 \pi ^{2}}}  + 
{ \frac {\sqrt{2}\,\eta }{\pi ^{4}}}  - 
{ \frac {\sqrt{2}\,\eta ^{5}}{\pi ^{4}}} \bigg)\,\mu ^{5/
2} \nonumber \\ && + \bigg( - { \frac {9
\sqrt{2}\,\eta }{128 \pi ^{2}}}
 \mbox{} + 
{ \frac {163 \sqrt{2}\,\eta ^{5}}{1920 \pi ^{2}}}  + 
{ \frac {7 \sqrt{2}\,
\eta ^{3}}{128 \pi ^{2}}}  -
{ \frac {59 \eta ^{7}\,\sqrt{2}}{36 \pi ^{4}}}  + 
{ \frac {1117 \eta 
^{7}\,\sqrt{2}}{13440 \pi ^{2}}} \nonumber \\ && - 
{ \frac {7 \eta ^{3}\,\sqrt{2}}{12 \pi ^{4}}}  - 
{ \frac {\sqrt{2}\,
\eta ^{5}}{4 \pi ^{4}}}  + { 
\frac {25 \sqrt{2}\,\eta }{36 \pi ^{4}}} \bigg)\mu ^{7/2}\mbox{} + {\cal O}(
\mu ^{9/2}).
\end{eqnarray}

Related to the fact that the first four $f_n$ can be calculated without making
use of the Jacobi inversion problem, Bardeen and Wagoner were able to determine
them in \cite{BW}. Taking into account that the relativistic parameter used in
said paper is somewhat different,~
\footnote{Bardeen and Wagoner chose to use the relativistic parameter $\gamma$
defined by $\gamma=1-e^{V_0}$. The series forms of the functions
$\gamma=\gamma(\mu)$ and $\mu=\mu(\gamma)$ can be calculated using the
function $V_0=V_0(\mu)$ (see \cite{NM94}) and are listed up to the orders
$\mathcal{O}(\mu^{12})$ and $\mathcal{O}(\gamma^{12})$ respectively in
\cite{Diplomarbeit}.}
the coefficients calculated by Bardeen and Wagoner were compared with
eqs.~(\ref{eq:f1inuvw})-(\ref{eq:f4inuvw}) and found to be in agreement. Using
the knowledge that these first coefficients are correct, a further twelve $f_n$
have been proved correct by showing that they satisfy the Ernst equation up to
the relevant order. The further eight coeffecients that were explicitly
calculated could not undergo the same test due to computer limitations, but
were calculated using the same iteration scheme, whence their correctness is
all but certain. For more details see \cite{Diplomarbeit}.

While a direct comparison of the PN expansion of the Ernst potential to the
numerical results of Bardeen and Wagoner is not feasible for an arbitrary point
in space, various quantities in the disc can be compared quite readily. In
particular, the analytical version of Table 2 in \cite{BW} was drawn up,
presenting the series for the square of the linear velocity, $v^2$, for some
particle of the disc as measured in the locally nonrotating frame of reference.
It turns out that the numerical results, which are listed to the $5^{\rm th}$
decimal place, are perfectly correct up to the order $\mathcal{O}(\gamma^5)$.
\footnote{The results concerning $v^2$ were calculated using the parameter
$\gamma$ in order to facilitate the comparison to the work of Bardeen and
Wagoner.}
As to be expected, the expansion coefficients differ more and more from the
analytic values as one moves to higher orders of the series. The last entry in
Table 2 of \cite{BW} deviates from the analytic result by approximately 25\%.
In all however, the numerical results in Bardeen and Wagoner are highly
accurate and seldom differ from the analytic ones by more than a fraction of a
percent.
\subsection{Convergence}
\label{sec:convergence}
\begin{figure}
\begin{center}
\epsfig{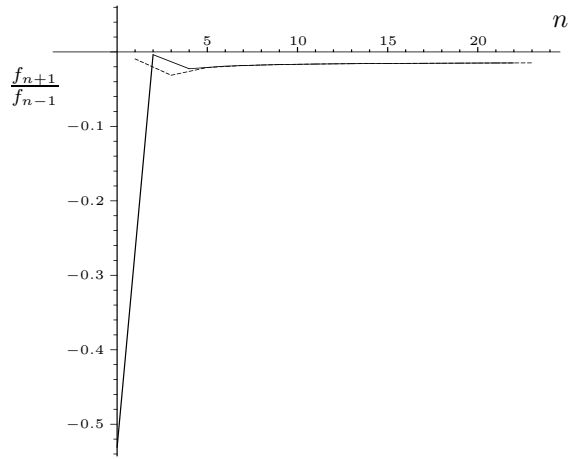}
\caption {The ratio of consecutive real and imaginary coefficients from the
expansion of $f$ taken at a point $\xi=0,\ \eta=3/5$ in the disc. The solid
line represents the real coefficients and the dotted line the imaginary
coefficients. \label{fig:disc}}
\end{center}
\end{figure}

As could have been expected right from the outset, the convergence of the PN
approximation depends to a great extent on position, i.e. on the coordinates
$\xi$ and $\eta$. For example, on the rim of the disc itself, the Ernst
potential is given by the analytic expression $f=1-\mu/2$, whereas in general,
the potential is given by an infinite series in $\mu$. Plots similar to that of
Fig.~\ref{fig:disc} strongly indicate that the series converges for spatial
infinity ($\xi \to \infty$) and has a radius of convergence of precisely
$\mu_0$ in this limit. Evaluations of the series at various (arbitrary) points
in space indicate, moreover, that it converges everywhere. The series tends to
converge quite quickly for small $|\eta|$, but rather erratically for $|\eta|$
nearing $1$ ($\eta=\pm 1$ corresponds to the axis of symmetry).

\begin{figure}
\begin{center}
\begin{overpic}[height=8.6cm,width=12cm,angle=270]{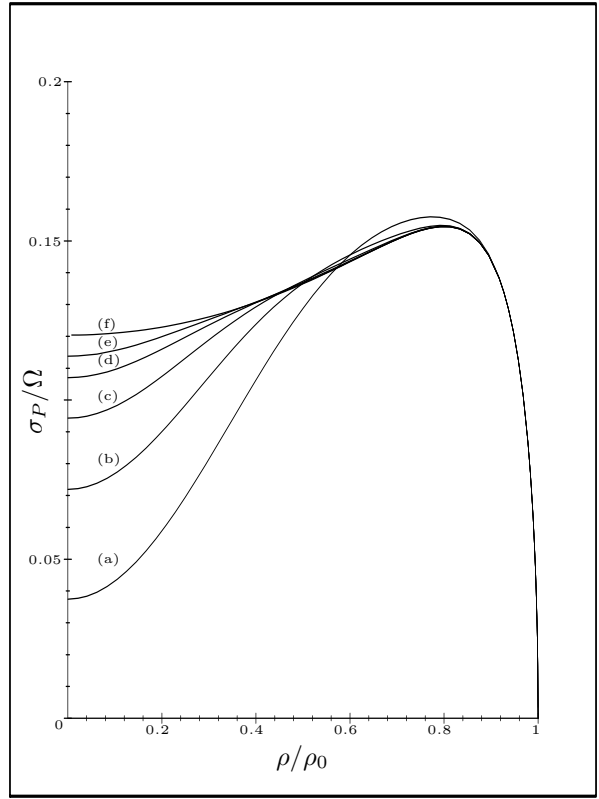}
\put(13,32){{\tiny (a)}}
\put(13,43){{\tiny (b)}}
\put(13,50){{\tiny (c)}}
\put(13,54){{\tiny (d)}}
\put(13,56){{\tiny (e)}}
\put(13,58){{\tiny (f)}}
\end{overpic}
\end{center}

\caption{The radial distribution of the proper surface mass density for $\mu=3$
at different orders of the PN approximation. The curves were created by
employing the approximation up to the order (a)~${\cal O}(\mu^{9/2})$,
(b)~${\cal O}(\mu^{13/2})$, (c)~${\cal O}(\mu^{17/2})$, (d)~${\cal
O}(\mu^{21/2})$ and (e)~${\cal O}(\mu^{25/2})$. The curve (f) was created by
evaluating the exact solution numerically to extremely high accuracy.
\label{fig:massendichte}}

\end{figure}

The Pad\'e approximant, which represents a truncated series via a rational
polynomial expression, proves to be a highly advantageous alternative to the
series, in particular if the original series converges slowly, erratically or
not at all. Using this approximant, the erratic convergence of the PN expansion
for the Ernst potential near the axis or the particularly slow convergence of
the dimensionless quantity $\Omega \rho_0$ can be largely circumvented.
Frequent use of the Pad\'e approximant was made in the work of Bardeen and
Wagoner \cite{BW}. In this paper however, sparing use of it is made so as
better to concentrate on the PN expansion itself, but a comparison of the
Pad\'e approximant to the standard PN series can be found in
Table~\ref{tab:pade}. Let it be mentioned that the proper surface mass density
(Fig.~\ref{fig:massendichte}) and the ergospheres (Figs.~\ref{fig:mu2} and
\ref{fig:mu3}) can be found to extremely high accuracy using the Pad\'e
approximant. A general discussion of the Pad\'e approximant can be found in
\cite{BO} and a consideration of its use in the PN expansion in \cite{DIS}.

In Figure~\ref{fig:disc}, the ratio of consecutive coefficients from the real
and imaginary parts of the Ernst potential has been plotted to illustrate the
convergence of the series for a chosen point in space. In order to have a
measure for the leading terms of the expansion, we took $f_0$ to be $1$ and
$f_{-1}$ to be $i$. Note that the series is guaranteed to converge at a given
value of $\mu$, say $\mu=\mu_c$, so long as the condition $\lim_{n \to \infty}
|f_{n+1}/f_{n-1}|<1/\mu_c$ is met. Thus the series always converges in the
Newtonian limit $\mu \to 0$ and converges in the limit $\mu \to \mu_0$ so long
as $\lim_{n \to \infty} |f_{n+1}/f_{n-1}| < 1/\mu_0 \approx 0.21$ holds.

In Fig. \ref{fig:disc} one sees that the series converges quite quickly for
this point
on the disc, 
that $f_1$ plays a fairly important role, but that $f_3$ and higher
coefficients contribute little to the potential. The fact that the ratio of
consecutive coefficients of the series is negative means that these
coefficients alternate sign, so that one always has an upper and lower bound on
the value for (the real and imaginary parts of) $f$. Note that not all points
in space exhibit such clear convergence as that of Fig.~\ref{fig:disc}.

Not only does the applicability of the series depend upon the point in space
that is chosen, but of course upon the value of $\mu$ as well. It goes without
saying that the PN approximation reflects the full field equations fairly
accurately for small values of $\mu$, but we shall see further, as did Bardeen
and Wagoner much to their surprise, that the approximation can be utilized to
great benefit even in many highly relativistic situations.
%
%
\begin{table}

\begin{center}
\begin{tabular}{|>{\PBS\raggedright}p{1.8cm}| 
               *{3}{>{\footnotesize\PBS\raggedright}p{2.5cm}|}} \hline
& $\mu=1/2$ & $\mu=3$  & $\mu=\mu_0$
\\ \hline

$\xi=1/2$, $\eta=1/2$
& $0.772687415-0.022438224i$
${\bf 0.772687415}-{\bf 0.022438224{\it i}}$
& $-0.281941980-0.352810399i$
${\bf -0.281941981}-{\bf 0.352810399{\it i}}$
& $-0.861419387-0.684924117i$
${\bf -0.861419898}-{\bf 0.684924396{\it i}}$
\\ \hline
On the axis $\xi=1$, $\eta=1$
& $0.820284295-0.032540154i$
${\bf 0.820284295}-{\bf 0.032540154{\it i}}$
& $0.127426794-0.560328316i$
${\bf 0.127409032}-{\bf 0.560315954{\it i}}$
& $0 - \hspace{3cm} i$
${\bf -0.006959726}-{\bf 0.997783811{\it i}}$
\\ \hline
In the disc $\xi=0$, $\eta=3/5$
& $0.675204587-0.041948129i$
${\bf 0.675204587}-{\bf 0.041948129{\it i}}$
& $-0.626305938-0.409718317i$
${\bf -0.626305488}-{\bf 0.409718090{\it i}}$
& $-1.304973284-0.616799888i$
${\bf -1.304859086} -{\bf 0.616728538{\it i}}$
\\ \hline
\end{tabular} \end{center}
\caption{The Ernst potential $f$ according to a numerical evaluation of the
exact solution in comparison with the results of the PN approximation
(bold-faced) up to the order $\mathcal{O}(\mu^{25/2})$ for various points in
space and three values of~$\mu$.
\label{tab:con}}
\end{table}
%
%
Looking at Table~\ref{tab:con}, one can see that the PN approximation of order
$\mathcal{O}(\mu^{25/2})$ returns values for $f$ that are correct to at least 9
decimal places at $\mu=1/2$ (the numerically evaluated potential, determined
using the analytic solution, is correct at least as far as it is given in
Table~\ref{tab:con}). The highly relativistic case $\mu=3$ can be handled more
than satisfactorily by the PN approximation, whereby the potential on the axis
is valid to only four decimal places. The accuracy of the series approximation
is astonishingly high even for the  limit $\mu \to \mu_0$.~
\footnote{It should be noted that $\rho_0$ tends to zero in this limit. The
extreme Kerr metric is obtained for $\rho^2+\zeta^2 \neq 0$, whereas the
analytic solution shows that finite values for $\xi$, as considered here, lead
to a different (not asymptotically flat) spacetime (see \cite{M98}). The PN
approximation, on the other hand, erroneously yields an asymptotically flat
spacetime in this limit, at least so long as the series is finite.}
Although the results on the axis are reasonably good in this limit (accurate to
within 1\%), one can obtain much more accurate results using the Pad\'e
approximant. Table~\ref{tab:pade} provides a comparison of the value of $f$ for
a point along the axis returned by the PN series to that of the ``diagonal''
Pad\'e approximant, whereby diagonal is meant to indicate the approximant for
which the polynomial in the numerator is of the same order as the polynomial in
the denominator. The values in this table should be compared to the ``exact''
value found in the cell of the $2^{{\rm nd}}$ column and $2^{{\rm nd}}$ row of
Table~\ref{tab:con}. It turns out that the values returned by the Pad\'e
approximant are so accurate as to provide a viable alternative to numerical
methods for all practical purposes.
\footnote{However, as $\mu \to \mu_0$, one has to be careful for large $\xi$.}
Using Table~\ref{tab:pade} one can see, moreover, that the nature of the
convergence of the Pad\'e approximation is more uniform and hence more
predictable than that of the PN approximation. Thus one could have come up with
a good estimate for the accuracy of the Pad\'e values in the table, but not for
the PN values. The PN value of order ${\cal O}(\mu^{21/2})$, for example, is
scarcely more accurate than that of order ${\cal O}(\mu^{13/2})$ whereas the
corresponding Pad\'e values have won a further 8 decimal places of accuracy. As
a rule of thumb, it appears one can rely on the Pad\'e approximant to gain one
decimal place of accuracy for each increase of $\sqrt{\mu}$ in the order of the
polynomial.

\begin{table}

\begin{center}
\begin{tabular}{|>{\PBS\raggedright}p{0.7cm}| 
               *{5}{>{\footnotesize\PBS\raggedright}p{1.6cm}|}} \hline
& $\mathcal{O}(\mu^{9/2})$ & $\mathcal{O}(\mu^{13/2})$  &
  $\mathcal{O}(\mu^{17/2})$ & $\mathcal{O}(\mu^{21/2})$ &
  $\mathcal{O}(\mu^{25/2})$
\\ \hline

PN
& $0.150938555 - 0.564116343i$
& $0.127710083 - 0.556936459i$
& $0.126549231 - 0.561010096i$
& $0.127627187 - 0.560281877i$
& $0.127409032 - 0.560315954i$
\\ \hline
Pad\'e
& $0.140716023 - 0.562426687i$
& $0.127381473 - 0.560314135i$
& $0.127426866 - 0.560328310i$
& $0.127426794 - 0.560328316i$
& $0.127426794 - 0.560328316i$
\\ \hline
\end{tabular} \end{center}

\caption{The Ernst potential $f$ for $\mu=3$ at the point $\xi=1$, $\eta=1$
according to the PN approximation and the Pad\'e approximant for different
orders of the series (cf. $2^{\rm nd}$ row, $2^{\rm nd}$ column of
the previous table).
\label{tab:pade}}
\end{table}

A pictorial impression of the convergence of the PN series can be gleaned from
Fig.~\ref{fig:massendichte}. The proper surface surface mass density
$\sigma_P$, which can be calculated from $\sigma_P=\sigma e^{U-k}$ where
$\sigma$ is given by eq.~(23) of \cite{NM94}, is divided by $\Omega$ to give a
dimensionless quantity and plotted as a function of the normalized radius
$\rho/\rho_0$. A discussion of the curve itself, which lies outside the
scope of this paper, can be found in \cite{BW}. What is of interest here is the
way in which the curves approach the curve (f). Although the tendency to
converge can be seen quite clearly, the discrepancy between curves (e) and (f)
is fairly large. This is primarily due to the slow convergence of the
dimensionless quantity $\Omega \rho_0$. For example, the PN approximation of
order ${\cal O}(\mu^{25/2})$ returns a value of approximately 0.220 instead of
0.213 for $\Omega \rho_0$ at $\mu=3$ and at $\mu=4.5$ the PN approximation
yields the grossly erroneous value of 1.46 as compared with the correct value,
$1.01 \times 10^{-7}$. With the Pad\'e approximant, this problem vanishes and
the curves for the proper surface mass density are indistinguishable from the
exact curves even at the order ${\cal O}(\mu^{9/2})$.

\psfrag{0.15}[r][r]{{\tiny$0.15$}}
\psfrag{0.1}[r][r]{}
\psfrag{0.05}[r][r]{{\tiny$0.05$}}
\psfrag{rhod}[t][t]{$\rho/\rho_0$}
\psfrag{y}[r][r]{\begin{sideways}{$\sigma_P/\Omega$}\end{sideways}}

\subsection{Ergospheres}
The results of Table~\ref{tab:con}, i.e. the fact that the PN approximation
returns a highly accurate value for $f$ even in very relativistic situations,
justify the use of this approximation even for the consideration of purely
relativistic phenomena. One such phenomenon, the ergosphere, dramatically
highlights differences between the Newtonian and Einsteinian theory. An
ergosphere is a region in which the Killing vector characterising the stationary
nature of the spacetime becomes spacelike (the metric function $e^{2U}$ is
negative in this region) so that nothing can remain still relative to an
observer at infinity --- everything is dragged along by the rotation of the
disc.

\begin{figure}
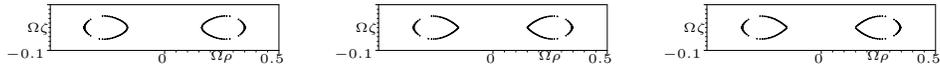
 \noindent
\begin{minipage}{.28\linewidth}
\centering\epsfig{figure=ergomu2or1201.eps,width=\linewidth}
\end{minipage}\hfill 
\begin{minipage}{.28\linewidth}
\centering\epsfig{figure=ergomu2or1601.eps,width=\linewidth}
\end{minipage}\hfill
\begin{minipage}{.28\linewidth}
\centering\epsfig{figure=ergomu2or2401.eps,width=\linewidth}
\end{minipage}\hfill
\caption{The ergospheres as calculated using the $6^{\rm
th}$, $8^{\rm th}$ and $12^{\rm th}$ PN approximation for $\mu=2$. The outline
represents the curve along which $e^{2U}$ is zero and should be
thought of rotated about the axis of symmetry. The inside of the
resulting torus-like figure of revolution is then the ergosphere.}
\label{fig:mu2} \end{figure}

\begin{figure}
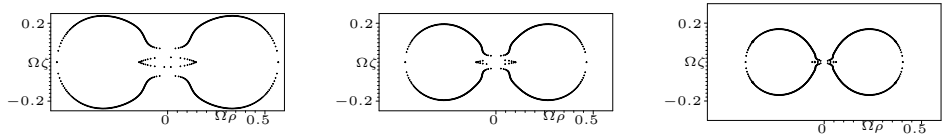
 \noindent
\begin{minipage}{.28\linewidth}
\centering\epsfig{figure=ergomu3or1201.eps,width=\linewidth}
\end{minipage}\hfill
\begin{minipage}{.28\linewidth}
\centering\epsfig{figure=ergomu3or1601.eps,width=\linewidth}
\end{minipage}\hfill
\begin{minipage}{.28\linewidth}
\centering\epsfig{figure=ergomu3or2401.eps,width=\linewidth}
\end{minipage}\hfill
\caption{The ergospheres as calculated using the $6^{\rm
th}$, $8^{\rm th}$ and $12^{\rm th}$ PN approximation for $\mu=3$. The outline
represents the curve along which $e^{2U}$ is zero and should be
thought of rotated about the axis of symmetry. The inside of the
resulting torus-like figure of revolution is then the ergosphere.}
\label{fig:mu3} \end{figure}

Figures~\ref{fig:mu2} and \ref{fig:mu3} depict the ergosphere for two values of
$\mu$ and calculated for three different orders of the PN approximation (please
see \cite{MK95} for a comparison with ergospheres found by evaluating the
analytic solution numerically). One can see in Fig.~\ref{fig:mu2} that the
ergosphere for $\mu=2$, as given by the $6^{\rm th}$ PN approximation, is nigh
on correct, so that higher order terms serve only to refine fine details. In
the case of $\mu=3$, however, (Fig.~\ref{fig:mu3}) one can see marked
improvement as one moves toward higher orders of the approximation.

As $\mu$ increases, the ergosphere moves closer and closer to the axis of
rotation where the PN approximation is no longer reliable. This, compounded
with the fact that the increasingly poor convergence of the dimensionless
quantity $\Omega \rho_0$ as $\mu$ nears $\mu_0$ results in scaling problems,
leads one to assume that the PN approximation is ill-suited for rendering the
ergoshpere for values of $\mu$ approaching $\mu_0$. Due to the extreme accuracy
of the Pad\'e approximant however, one can indeed divine the nature of the
ergospheres even in this limit.
%
%
\psfrag{ox}[t][t]{\tiny{$\Omega \rho$}}
\psfrag{oy}[r][r]{\tiny{$\Omega \zeta$}} 

\section{Conclusion}
It was possible, due to the existence of an analytic, global solution for the
axially symmetric, stationary, rigidly rotating disc of dust, to come up with
an iteration scheme to calculate an arbitrary coefficient in the PN
approximation of this solution. This work amounts to the analytic analogue of
numerical work published by Bardeen and Wagoner in 1971. The explicit
calculation of the expansion coefficients of the Ernst potential was carried
out to the $12^{\rm th}$ PN level (i.e. $\mathcal{O}(\mu^{25/2})$). It turns out
that the numerical results from Bardeen and Wagoner are highly accurate for
lower orders of the expansion and remain quite good even for higher orders.

The supposition that the PN approximation could be used for many highly
relativistic situations was confirmed for the disc of dust, and the very
accurate renderings of the ergospheres obtained using the PN approximation
attest to the fact that physically meaningful, relativistic phenomena can be
studied in some cases using tools tailored to other applications (i.e. to the
consideration of the Newtonian regime). It turns out that the position in space
to be considered contributes significantly to the convergence of the PN
approximation, but not to the convergence of the corresponding Pad\'e
approximant. It was found that the PN approximation is unreliable for large
$\mu$ in the neighbourhood of the axis of rotation even when the approximation
is quite accurate elsewhere for the same value of $\mu$, and  it can be made
highly accurate for all points in space at all values of $\mu$ by applying the
Pad\'e approximation methods. If one is careful to take  into account its
limitations, one can use the PN approximation to great advantage for quick and
accurate calculations and is bolstered in the opinion that approximation
methods can sometimes be extended to applications outside their {\it a priori}
guaranteed region of validity.

\section*{}

\end{document}